\documentclass[10pt,leqno]{amsart}
\usepackage{amsmath,amsthm}

\def\init{\setcounter{equation}{0}}
\newcommand{\e}{\varepsilon}
\newcommand{\rw}{\rightarrow}
\usepackage{tikz}
\usetikzlibrary{decorations.pathreplacing}

\begin{document}
\newtheorem{theorem}{Theorem}[section]
\newtheorem{corollary}{Corollary}[section]

\newcommand{\R}{\mathbb{R}}
\newcommand{\C}{\mathbb{C}}
\newcommand{\Z}{\mathbb{Z}}

\newcommand{\integ}{\int_{\R^2}}
\newcommand{\h}{\stackrel{\rightarrow}{h}}
\newtheorem{lemma}{Lemma}[section]
\newtheorem{pro}{Proposition}[section]
\newcommand{\Tr}{\operatorname{Tr}}
\newcommand{\D}{\mathcal{D}}


\title{New examples of Hawking radiation  from   acoustic black holes}
\author{Gregory Eskin}


\maketitle

In memory  of Selim Grigorievich Krein 
 on occasion of his 100th birthday
\\ 
\ 
\\
\begin{abstract}
Rotating acoustic metrics  may have black holes 	inside the ergosphere.  Simple  cases of acoustic black holes were studied	
in the references  [21],  [4]. 
 In the present paper we study  the Hawking radiation for
 more complicated  cases  of acoustic black holes
including black holes  that have corner points.																																																																																																																											
\end{abstract}

\section{Introduction.}
\init

In classical general relativity black hole
is the region  such that no signal or disturbance can escape it. 
 It was  a remarkable discovery  by S. Hawking  [10] that once the quantum effects are added  the black 
holes emit  quantum  particles.  This effect  is called  the Hawking radiation and it was studied subsequently in many papers
(cf.  [1],  [2],  [3],  [8],  [9],  [12],  [13],  [14]   and others).   In an  astrophysical  
experiment the  Hawking effect is too weak to be detected on the earth.

In quest to find the experimental confirmation of the Hawking radiation effect   W. Unruh considered  the Hawking radiation 
for the acoustic black holes  [18].   Acoustic black holes are one of the examples  of analogue black holes  (cf.  [21],  see also  [15],  [16]).   The
Hawking radiation  for analogue  black  hole have been studied  in [19],  [20],  [22]  and others.

In all previous works  on Hawking  radiation  the case of spherically symmetric metric or the case  of one space  
dimension  was treated.

In present paper,  as in [6],  we study the case of two space dimensional (rotating)  acoustic metric.  It was shown  in 
[4]  and [5]  that there is a rich class  of  acoustic  metrics  having black holes.   In  [6]  we consider the case of smooth
black holes  but as it was shown in [5]  there exist   black  holes that are not smooth     
and that event horizons may have corners.

The Hawking radiation  from such black  holes  will be considered in this paper.

The plan of the paper is the following:

In \S 2   we set up the framework  of the quantum field  theory  on curved spacetimes   
following mostly  T. Jacobson  [12],  [13]  (see also  [8], [11]).  In departure  from  [6],  where  the  Unruh type vacuum [17]    was considered,
we use in \S 2 a more common  vacuum state  as  in S. Hawking  [10]  and  T. Jacobson  [12].

In \S   3 we  briefly  
consider   the case   of a simple acoustic black  hole  similar   to the one   in \S 3  of  [6].   There  is a difference  in computations
with \S 3  of [6]  since  we consider  a different vacuum state  in this paper.

In \S4  we study  the Hawking radiation  for the acoustic  metrics when  the ergosphere  has  a finite  number  
of characteristic points.  This case is different  from  the case considered in section 4.1 of [6]  where the characteristic points were 
not allowed.  In \S 5  we consider  the Hawking radiation  from a black  holes having corners.
Note that  constructions  in \S4 and \S 5 requires a localization in the  angular  coordinate.

\section{The number of particles operator}
\init

Consider a fluid flow  in a vortex  with the velocity field
\begin{equation}														\label{eq:2.1}
v=(v^1,v^2)=\frac{A}{r}\hat x+\frac{B}{r}\hat\theta,
\end{equation}
where 
$r=|x|,\ \hat x=\frac{(x_1,x_2)}{|x|},\ \hat\theta= \frac{(-x_2,x_1)}{|x|},\ A,B$  are functions of $x=(x_1,x_2)$.

The acoustic waves in the moving fluid  are described by the wave equation  (cf.  [21])
\begin{equation}                                          								       \label{eq:2.2}
\Box_g u(x_0,x_1,x_2) =\frac{1}{\sqrt{g(x_1,x_2)}}\sum_{j,k=0}^2\frac{\partial}{\partial x_k}\left(\sqrt{g(x_1,x_2)}g^{jk}(x_1,x_2)
\frac{\partial u(x_0,x_1,x_2)}{\partial x_j}\right)=0,
\end{equation}
$x_0$  is the time variable,  $g^{00}=1,\ g^{0j}=g^{j0}=v^j,\  g^{jk}=(-\delta_{jk}+v^jv^k),
\linebreak
1\leq j,k\leq 2,\   g(x_1,x_2)=1$  and we,  for the simplicity,  assume  the sound  speed and density are  equal  to 1.

The Hamiltonian corresponding  to   (\ref{eq:2.2})   has the following  form  in polar  coordinates  $(\rho,\varphi)$:
\begin{equation}															\label{eq:2.3}
H(\rho,\varphi,\eta_0,\eta_\rho,\eta_\varphi)=\Big(\eta_0+\frac{A}{\rho}\eta_\rho +\frac{B}{\rho^2}\eta_\varphi\Big)^2
-\eta_\rho^2-\frac{1}{\rho^2} \eta_\varphi^2,
\end{equation}
where $(\eta_0,\eta_\rho,\eta_\varphi)$  are dual to  $(x_0,\rho,\varphi)$.

Note that  (\ref{eq:2.3})  can  be  factored  
$$
H=H^+H^-,
$$
where 
\begin{equation}																\label{eq:2.4}
H^\pm(\rho,\varphi,\eta_0,\eta_\rho,\eta_\varphi)=
\eta_0+\frac{A}{\rho}\eta_\rho +\frac{B}{\rho^2}\eta_\varphi \pm
\sqrt{\eta_\rho^2 +\frac{1}{\rho^2}\eta_\varphi^2}.
\end{equation}
We refer  to [11],  [12]  to introduce  the main facts of quantum  field theory  on curved spacetimes.

  We define
  $f_k^+(x_0,x)$  as the solution  of  (\ref{eq:2.2})  in  $\R^2\times\R$   having  the following  initial  conditions in polar coordinates 
$(\rho,\varphi)$
\begin{align}																	\label{eq:2.5}
&f_k^+(x_0,x_1,x_2)\Big|_{x_0=0}=\gamma_k e^{ik\cdot x},
\\
\nonumber
&\frac{\partial}{\partial x_0}f_k^+(x_0,x_1,x_2)\Big|_{x_0=0}=i\lambda_0^-(k)\gamma_ke^{ik\cdot x},
\end{align}
where
$x=(\rho,\varphi), \ k=(\eta_\rho,m),\ m\in\Z,
\ k\cdot x=\rho\eta_\rho+m\varphi,
\ \gamma_k=\frac{1}{2\pi \sqrt 2\sqrt \rho\big(\eta_\rho^2+a^2\big)^{\frac{1}{4}}},\linebreak a>0$  is arbitrary,
\begin{equation}																\label{eq:2.6}
\lambda_0^-(k)=-\frac{A}{\rho}\eta_\rho  -\frac{Bm}{\rho^2}-\sqrt{\eta_\rho^2+a^2}.
\end{equation}
 Let
\begin{equation}																	\label{eq:2.7}
   f_k^-(x_0,x)= \overline{f_{-k}^+(x_0,x)}.
\end{equation}
Let $u,v$  be the solutions of  (\ref{eq:2.2})   and  $<u,v>  $   be  
their Klein-Gordon  inner product
\begin{equation}																\label{eq:2.8}
<u,v>=i\int\limits_{x_0=t} |g|^{\frac{1}{2}}\{u,v\}dx_1 dx_2,
\end{equation}
where $\{u,v\}=\sum_{j=0}^2  g^{0j}\Big(\overline u\frac{\partial v}{\partial x_j}-\frac{\partial\overline u}{\partial x_j}v\Big)$.

Note that (\ref{eq:2.8})  is independent   of $t$  (cf.  [12]).

It is easy to see (cf. [6])  that
\begin{equation}																\label{eq:2.9}
<f_k^+,f_{k'}^+>=\delta(\eta_\rho-\eta_\rho')\delta_{mm'},\ \ <f_k^-,f_{k'}^->=-\delta(\eta_\rho-\eta_\rho')\delta_{mm'},
\ \ <f_k^+,f_{k'}^->=0,
\end{equation}
where   $k=(\eta_\rho,m),k'=(\eta_\rho',m')$.

Thus $f_k^+,f_{-k}^-$  form a basis of solutions of (\ref{eq:2.2}).

Now we introduce  the selfadjoint field operator $\Phi$
\begin{equation}																\label{eq:2.10}
\Phi=\sum_{m=-\infty}^\infty \int\limits_{-\infty}^\infty \big (\alpha_{k}^+f_{k}^+(x_0,x_1,x_2)+\alpha_{-k}^-f_{-k}^-(x_0,x_1,x_2)\big )d\eta_\rho,
\end{equation}
where
$k=(\eta_\rho,m),\eta_\rho\in \R,m\in\Z $,
\begin{equation}																\label{eq:2.11}
\alpha_k^+=<f_k^+,\Phi>,\ \ \alpha_{k}^-=- <f_{k}^-,\Phi>.
\end{equation}
Since  $\overline f_k^+=f_{-k}^-$  we have   that  $(\alpha_{-k}^-)^*=\alpha_k^+$.

Operators
$\alpha_k^+,\alpha_{-k}^-$   are called  the  annihilation and creation  operators,  respectively,  and they satisfy  
the following commutation relations (cf. [12]):
\begin{equation}																\label{eq:2.12}
[\alpha_k^+,\alpha_{-k'}^-]=\delta(\eta_\rho-\eta_\rho')\delta_{mm'} I,\ \ [\alpha_k^+,\alpha_{k'}^+]=0,
\end{equation}
  $[\alpha_{k}^-,\alpha_{k'}^-]=0,\ I$  is the identity  operator.

Let  $C(x_0,x_1,x_2)$ be a solution  of  (\ref{eq:2.2})  with
some
 initial conditions at  $x_0=0$   that will be specified later.

  Expanding  $C(x_0,x_1,x_2)$  
in  the basis  $f_k^+,f_{-k}^-$   we get
\begin{equation}																\label{eq:2.13}
C=\sum_{m=-\infty}^\infty\int\limits_{-\infty}^\infty\big( C^+(k)f_k^+(x_0,x_1,x_2)
+ C^-(k)f_{-k}^-(x_0,x_1,x_2)\big) d\eta_\rho,
\end{equation}
where  $k=(\eta_\rho,m),\ C^+(k)=<f_k^+,C>,\ C^-(k)=- <f_{-k}^-,C>$.

Let
\begin{equation}																\label{eq:2.14}
C^\pm=\sum_{m=-\infty}^\infty\int\limits_{-\infty}^\infty C^\pm(k)f_{\pm k}^\pm(x_0,x) d\eta_\rho,
\end{equation}

Thus 
\begin{equation}     																\label{eq:2.15}
C=C^+  + C^-.
\end{equation}

It follows from  (\ref{eq:2.10}),  (\ref{eq:2.13})   that 
\begin{equation}																\label{eq:2.16}
<C,\Phi>=\sum_{m=-\infty}^\infty\int\limits_{-\infty}^\infty\big( \overline {C^{+}}(k)\alpha_k^{+}
-\overline {C^{-}}(k)\alpha_{-k}^{-}
\big) d\eta_\rho.
\end{equation}

 The vacuum  state $| 0\rangle$  is defined
by the conditions
\begin{equation}														\label{eq:2.17}
\alpha_k^{+}| 0\rangle  =0\ \ \mbox{for all\ }  k.
\end{equation}

Let   $N(C)$  be  the number of particles  operator  created  by  the wave  packet $C$ (cf.  [12])
\begin{equation}														\label{eq:2.18}
N(C)=<C,\Phi>^*<C,\Phi>,
\end{equation}
and let  $\langle 0| N(C) |0 \rangle$   be the average number of  particles.

As in [6]  one have the following theorem
\begin{theorem}														\label{theo:2.1}
The average  number of particles  created  by the wave packet  $C$  is given by the formula
\begin{equation}														\label{eq:2.19}
\langle 0| N(C) | 0\rangle=\sum_{m=-\infty}^\infty\int\limits_{-\infty}^\infty| C^{-}(k)|^2 d\eta_\rho,
\end{equation}
where $C^{-}$  is the same  as in (\ref{eq:2.14}).
\end{theorem}
Note that  
(cf.  (\ref{eq:2.14}))  
\begin{equation}  															\label{eq:2.20}
<C^-,C^->=-\sum_{m=-\infty}^\infty\int\limits_{-\infty}^\infty|C^-(k)|^2d\eta_\rho.
\end{equation}
Therefore
\begin{equation}															\label{eq:2.21}
\langle 0 |N(C)|0\rangle=-<C^-,C^->,
\end{equation}
where                                                                                                                                                                                                                                                                                                                                                                                                                                                                                                                                                                                                                                                                                                                                                                                                                                                                                                                                                                                                                                                                                                                                                                                                                                                                                                                                                                                                                                                                                                                                                                                                                                                                                                                                                                                                                                                                                                                                                                                                                                                                                                                                                                                                                                                                                                                                                                                                                                                                                                                                                                                                                                                                                                                                                                                                                                                                                                                                                                                                                                                                                                                                                                                                                                                                                                                                                                                                                                                                                                                                                                                                                                                                                                                                                                                                                                                                                                                                                                                                                                                                                                                                                                                                                                                                                                                                                                                                                                                                                                                                                                                                                                                                                                                                                                                                                                                                                                                                                                                                                                                                                                                                                                                                                                                                                                                                                                                                                                                                                                                                                                                                                                                                                                                                                                                                                                                                                                                                                                                                                                                                                                   
$C^-$  is the same as in  (\ref{eq:2.14}).
\\
{\bf Proof of Theorem  2.1}
We have
\begin{equation}															\label{eq:2.22}
<C,\Phi > |0\rangle =-\sum_{m=-\infty}^\infty\int\limits_{-\infty}^\infty\overline{C^-(k)}\alpha_{-k}^- | 0\rangle d\eta_\rho.
\end{equation}
Analogously,
\begin{equation}															\label{eq:2.23}
\langle 0|<C,\Phi>^* =-\sum_{m=-\infty}^\infty\int\limits_{-\infty}^\infty  \langle 0|C^-(k)\alpha_{k}^+  d\eta_\rho,
\end{equation}
since  $\langle 0|\alpha_{-k}^-=0$.
Therefore
\begin{multline}															\label{eq:2.24}
\langle 0|<C,\Phi>^*  <C,\Phi > |0\rangle  
\\
=\sum_{m=-\infty}^\infty\int\limits_{-\infty}^\infty  \langle 0|C^-(k)\alpha_k^+  d\eta_\rho\cdot
\sum_{m'=-\infty}^\infty\int\limits_{-\infty}^\infty\overline{C^-(k')}\alpha_{-k'}^- | 0\rangle d\eta'_\rho
\\
=\sum_{m=-\infty}^\infty\int\limits_{-\infty}^\infty|C^-(k)|^2d\eta_\rho,
\end{multline}
since  
$\alpha_k^+\alpha_{-k'}^-=\alpha_{-k}^-\alpha_k^++I\delta(k-k')$  and   $\alpha_k^+|0\rangle=0,\ \forall k$.

\section{Hawking radiation  from the simple rotating acoustic black hole}
\init
Consider the simplest case when $A<0,  B\neq 0$  are constants.  This case was studied in [6], \S3,  and we briefly repeat the computations
taking into   account that the vacuum state  in  this  paper  is different  from  [6],  \S 3.

Let 
$S(x_0,\rho,\varphi)$   be the eikonal  satisfying the equation
$$
\Big(-\eta_0  +\frac{A}{\rho}S_\rho  +\frac{B}{\rho^2}S_\varphi\Big)^2-S_\rho^2-\frac{1}{\rho^2}S_\varphi^2=0.
$$
We  are looking  for the eikonal  of the form
$S= -\eta_0x_0+s(\rho)+m\varphi$  where  $s(\rho)\rw +\infty  $   when  $\rho > |A|$  and   $\rho\rw |A|$.

It can  be shown  (cf.  [6])  that when  $\rho \rw |A|$ 
$$
S_\rho=\frac{|A|\big(\eta_0-\frac{Bm}{|A|^2}\big)}{\rho-|A|}+O(\rho-|A|).
$$
We will not try to solve  the eikonal  equation   exactly.  Instead  we use  the approximation  
$$
S_1=-\eta_0x_0+|A|\Big(\eta_0-\frac{Bm}{|A|^2}\Big)\ln(\rho-|A|)+m\varphi
$$  
of  $S(x_0,\rho,\varphi)$ to construct  the exact  solution  of
(\ref{eq:2.2})  having the following initial conditions

\begin{align}																\label{eq:3.1}
&C(x_0,\rho,\varphi)\Big|_{x_0=0}=\theta (\rho-|A|)
\frac{(\rho-|A|)^\e}{\sqrt\rho}\  e^{-a(\rho-|A|)}\,  e^{i\xi_0|A|\ln(\rho-|A|)+im\varphi},
\\	
\nonumber															
&\frac{\partial C(x_0,\rho,\varphi)}{\partial x_0}\Big|_{x_0=0}=i\beta C(x_0,\rho,\varphi_0)\Big|_{x_0=0},
\end{align}
where 
$
a>0,\e >0$   are arbitrary,  $\theta(\rho-|A|)=1 \ \ \mbox{for}\ \   \rho>|A|\ \ \mbox{and}
 \ \  \  \theta(\rho-|A|)=0$  for  $\rho<|A|,\ \xi_0=\eta_0-\frac{B m}{|A|^2},
$
\begin{equation}																\label{eq:3.2}
\beta=-\frac{A}{\rho}\frac{\xi_0|A|}{\rho-|A|}-\frac{Bm}{\rho^2}-\frac{\xi_0|A|}{\rho-|A|}.
\end{equation}
For  the convenience we take $a>0$  in (\ref{eq:3.2})  equal  to $a$  in  (\ref{eq:2.6}).   It is easy  to compute    the KG norm of  
$C(x_0,\rho,\varphi)$  
(cf.    (3.8)  in [6]):
\begin{equation}  																\label{eq:3.3}
<C,C>=\frac{4\pi \Gamma(2\e)\xi_0|A|}{(2a)^\e}.
\end{equation}
We shall call $C(x_0,\rho,\varphi)$  a wave  packet.

By Theorem \ref{theo:2.1}
to compute the average  number  of created particles  we need to find $C^-(k)=- <f_{-k}^-,C>$.  Computing the  KG  norm  we have 
\begin{equation} 																	\label{eq:3.4}
C^-=C_1^-(k)+C_2^-(k),
\end{equation}
where  (cf.  (3.11),  (3.12)  in  [6])
\begin{multline}																\label{eq:3.5}
C_1^-(k)
=\int\limits_0^\infty\int\limits_0^{2\pi}\frac{e^{i\rho\eta_\rho+im'\varphi}\theta(\rho-|A|)}
{\sqrt\rho (\eta_\rho^2+a^2)^{\frac{1}{4}}\sqrt 2 \ 2\pi}
\ \frac{(\rho-|A|)^\e}{\sqrt \rho}\ 
\Big(\frac{\xi_0|A|}{\rho-|A|}-\frac{A}{\rho}\Big(\frac{-i\e}{\rho-|A|}+ia\Big)\Big)
\\
\cdot e^{-a(\rho-|A|)+i\xi_0|A|\ln(\rho-|A|)}\
  e^{im\varphi} \rho\, d\rho d\varphi,
\end{multline}
\begin{multline}																\label{eq:3.6}
C_2^-(k)
=-\int\limits_0^\infty\int\limits_0^{2\pi}\frac{e^{i\rho\eta_\rho+im'\varphi}(\eta_\rho^2+a^2)^{\frac{1}{4}}
}
{\sqrt\rho \sqrt 2 \ 2\pi}
\ \frac{\theta(\rho-|A|)(\rho-|A|)^\e}{\sqrt \rho}\ 
\\
\cdot e^{-a(\rho-|A|)+i\xi_0|A|\ln(\rho-|A|)+im\varphi}\
 \rho d\rho d\varphi,\ \ \ k=(\eta_\rho,m').
\end{multline}
Integrating  in  $\varphi$  and using the  formula  (cf.  [6])
\begin{equation}															\label{eq:3.7}
\int\limits_0^\infty  e^{it\eta_\rho}t^\lambda e^{-at}dt  =\frac{e^{i\frac{\pi}{2}(\lambda+1)}\Gamma(\lambda+1)}
{(\eta_\rho+ia)^{\lambda+1}},
\end{equation}
we get
\begin{multline}															\label{eq:3.8}
C_1^-(k)=\frac{\delta_{m',-m}}{\sqrt 2}e^{i|A|\eta_\rho}
\frac{e^{i\frac{\pi}{2}(i\xi_0|A|+\e)}\Gamma(i\xi_0|A|+\e)(\xi_0|A|-i\e)}
{(\eta_\rho^2+a^2)^{\frac{1}{4}}(\eta_\rho+ia)^{i\xi_0|A|+\e}} \Big(1-\frac{ia}{\eta_\rho+ia}\Big)
\\
 +\delta_{m',-m}O\Big(\frac{1}{|\eta_\rho+ia|^{1+\e}}\Big),
\end{multline}
\begin{equation}															\label{eq:3.9}
C_2^-(k)=-\frac{\delta_{m',-m}}{\sqrt 2}e^{i|A|\eta_\rho}
\frac{e^{i\frac{\pi}{2}(i\xi_0|A|+\e+1)}\Gamma(i\xi_0|A|+\e+1)(\eta_\rho^2+a^2)^{\frac{1}{4}}
}
{
(\eta_\rho+ia)^{i\xi_0|A|+\e+1}},
\end{equation}
where $\delta_{m_1,m_2}=1$  when  $m_1=m_2$  and  $\delta_{m_1,m_2}=0$  when   $m_1\neq m_2$.                                                                                                                                                                                                                                                                                                                                                                                                                                                                                                                                                                                                                                                                                                                                                                                                                                                                                                                                                                                                                                                                                                                                                                                                                                                                                                                                                                                                                                                                                                                                                                                                                                                                                                                                                                                                                                                                                                                                                                                                                                                                                                                                                                                                                                                                                                                                                                                                                                                                                                                                                                                                                                                                                                                                                                                                                                                                                                                                                                                                                                                                                                                                                                                                                                                                                                                                                                                                                                                                                                                                                                                        
Since  $\big|e^{i\frac{\pi}{2}(i\xi_0|A|+\e)}\big|
=e^{-\frac{\pi}{2}\xi_0|A|},\ \Gamma(i\xi_0|A|+\e+1)=(i\xi_0|A|+\e)e^{-\frac{\pi}{2}\xi_0|A|}\Gamma_1(i\xi_0|A|+\e),\linebreak \Gamma_1$  is bounded,
we have
\begin{multline}															\label{eq:3.10}
|C^-|^2=|C_1^- + C_2^-|^2=
\frac{\delta_{m',m}}{2}
\Big|\frac{\xi_0|A|-i\e}{(\eta_\rho^2+a^2)^{\frac{1}{4}}}\cdot\frac{\eta_\rho}{\eta_\rho+ia}
-\frac{i(i\xi_0|A|+\e)(\eta_\rho^2+a^2)^{\frac{1}{4}}}{\eta_\rho+ia}\Big|^2
\\
\cdot e^{-2\pi\xi_0|A|}
|\Gamma_1(i\xi_0|A|+\e)|^2e^{2\xi_0|A|\arg(\eta_\rho+ia)}
(\eta_\rho^2+a^2)^{-\e}
\\
+\delta_{m',-m}O\big(|\eta_\rho+ia|^{-2-2\e}\big).
\end{multline}
Taking  the    sum in $m'$,
integrating  in $\eta_\rho$  and   making  change  of variables $\eta_\rho\rw a\eta_\rho$,   we  get
\begin{equation}                       											\label{eq:3.11}
\langle 0|N(C)|0\rangle =a^{-2\e}\int\limits_{-\infty}^\infty C_3(\eta_\rho)d\eta_\rho +O\big(a^{-2\e-1}\big),
\end{equation}
where
\begin{multline}														\label{eq:3.12}
C_3(\eta_\rho)=\frac{1}{2}e^{-2\pi\xi_0|A|}
|\Gamma_1(i\xi_0|A|+\e)|^2\, |\xi_0|A|+i\e|^2
\ \Big|\frac{\eta_\rho}{(\eta_\rho^2+1)^{\frac{1}{4}}}+(\eta_\rho^2+1)^{\frac{1}{4}}\Big|^2
\\
\cdot e^{2\xi_0|A|\arg(\eta_\rho+i)}(\eta_\rho^2+1)^{-\e-1}.
\end{multline}
Denote  by  $C_n$   the normalized  wave  packet,  i.e.   $C_n=\frac{C}{<C,C>}$.    Thus   $N(C_n)=\frac{N(C)}{<C,C>}$.   Dividing   
(\ref{eq:3.11})   by   $< C,C>$  and  taking the limit  as  $a\rw \infty$  we  get
\begin{equation}													\label{eq:3.13}
\lim_{a\rw \infty} \langle 0| N(C_n)| 0\rangle=
\frac{2^\e }{4\pi \Gamma(2\e)}\int\limits_{-\infty}^\infty\frac{1}{\xi_0|A|}C_3(\eta_\rho)d\eta_\rho.
\end{equation}
Therefore  we proved the following theorem:
\begin{theorem}													\label{theo:3.1}
The average  number   of particles   created  by the normalized wave packet  $C_n$   is given  by (\ref{eq:3.13})  where  
$C_3(\eta_\rho)$  is  the same  as  in (\ref{eq:3.12}).
\end{theorem}

Now  we  will  analyze    the decay  of the  right size  of  (\ref{eq:3.13})  when  $\xi_0|A|\rw\infty$.
Note
\begin{equation}															\label{eq:3.14}
\arg (\eta_\rho+i)=\sin^{-1}\frac{1}{\sqrt{\eta_\rho^2+1}}\ \ \ \ \mbox{when}\ \ \ \eta_\rho>0,
\end{equation}
\begin{equation}															\label{eq:3.15}
\arg (\eta_\rho+i)=\pi - \sin^{-1}\frac{1}{\sqrt{\eta_\rho^2+1}}\ \ \ \ \mbox{when}\ \ \ \eta_\rho<0.
\end{equation}
We have
\begin{equation}															\label{eq:3.16}
\Big| \frac{1}{(\eta_\rho^2+1)^{\frac{1}{4}}}  +\frac{(\eta_\rho^2+1)^{\frac{1}{4}}} {\eta_\rho+i}\Big|^2
\end{equation}
\begin{equation} 															\label{eq:3.17}
\mbox{is }\ \ \ O\Big(\frac{1}{(\eta_\rho^2+1)^{\frac{1}{2}}}\Big)\ \ \ \mbox{when}\ \ \ \eta_\rho>0,
\end{equation}
\begin{equation} 															\label{eq:3.18}
\mbox{and }\ \ \ O\Big(\frac{1}{\eta_\rho^2+1}\Big) \ \ \ \mbox{when}\ \ \ \eta_\rho<0,
\end{equation}
Note that 
$ \sin^{-1}\frac{1}{\sqrt{\eta_\rho^2+1}}\leq \frac{\pi}{2}$.     Therefore for  $\eta_\rho>0$  we have  
\begin{equation}															\label{eq:3.19}
C_3\leq C  \frac{e^{-2\pi\xi_0|A|+\pi \xi_0|A|}((\xi_0|A|)^2+\e^2)}
{(\eta_\rho^2+1)^{\e+\frac{1}{2}}} ,\ \ \eta_\rho>0,
\end{equation}
i.e. $C_3$  is exponentially  decaying on $(0,+\infty)$.

When $\eta_\rho<0$  we have
\begin{equation}															\label{eq:3.20}
C_3\leq C \  
\frac{(\xi_0|A|)^2e^{-2\xi_0|A|\sin^{-1}\frac{1}{\sqrt{\eta_\rho^2+1}}}}{(\eta_\rho^2+1)^{\e+1}},\ \ \eta_\rho<0.
\end{equation}
If  $|\eta_\rho|\leq (\xi_0|A|)^\delta,\ 0<\delta<1$,  then
\begin{equation}															\label{eq:3.21}
e^{-2\xi_0|A|\sin^{-1}\frac{1}{\sqrt{\eta_\rho^2+1}}  }\leq e^{-C(\xi_0|A|)^{1-\delta}}.
\end{equation}
Thus $C_3$  is exponentially decaying on  $(-(\xi_0|A|)^\delta,0)$.
On  $(-\infty,-(\xi_0|A|)^\delta)$  we get 
\begin{multline}																\label{eq:3.22}
\int\limits_{-\infty}^{-(\xi_0|A|)^\delta} C_3\ d\eta_\rho
 \leq  \int\limits_{-\infty}^{-(\xi_0|A|)^\delta}C\frac{(\xi_0|A|)^2}{(\eta_\rho^2+1)^{\e+1}}d\eta_\rho
 \leq  C\frac{(\xi_0|A|)^2}{((\xi_0|A|)^{2\delta}+1)^{\e+\frac{1}{2}}}
\end{multline}
Hence
\begin{equation}															\label{eq:3.23}
\int\limits_{-\infty}^0 C_3\ d\eta_\rho\leq C(\xi_0|A|)^{2-\delta(1+2\e)}.
\end{equation}
  Therefore  normalizing $C$  to make  $< C_n, C_n>=1$,  we get  (cf. (\ref{eq:3.3}))
\begin{equation}															\label{eq:3.24}
\int\limits_{-\infty}^0\frac{1}{\xi_0|A|} C_3 d\eta_\rho\leq  c_1(\xi_0|A|)^{1-\delta(1+2\e)}.
\end{equation}   

{\bf Remark 3.1}  It follows  from  (\ref{eq:3.19})   that  
$\int_0^\infty C_3\, d\eta_\rho$  is  exponentially  decaying  in  $\xi_0|A|$   (cf.  [10]) and so  it contributes to the Hawking radiation.
   However,  the integral  $\int_{-\infty}^0 C_3\, d\eta_\rho$
has  only  decay of order  (\ref{eq:3.24})  and it does not contribute   to the Hawking radiation.
Note that  for  $f_k^+(x_0,x_1,x_2)$  and $f_{-k}^-(x_0,x_1,x_2)=\overline{f_k^*}$  to form a basis of solutions  of 
(\ref{eq:2.2})  one needs to use all $\eta_{\rho}\in (-\infty,+\infty)$.   Therefore  
$\langle 0|N(C)|0\rangle$  must  include  also
$\int_{-\infty}^0 C_3\, d\eta_\rho$.    Only when we replace  $|0\rangle$  by  the  Unruh  type vacuum  $|\Psi\rangle$  (cf. [6])  we  get  that all
terms  of  $\langle \Psi| N(C) |\Psi\rangle$   contribute  to the  Hawking radiation.

\section{Hawking radiation  from rotating acoustic black holes}
\init

In this and the next sections we  shall continue to study the Hawking  radiation from  rotating acoustic black 
holes started in [6].  We shall consider  the case of the fluid flow 
(\ref{eq:2.1})   where   $A(\rho,\varphi), B(\rho,\varphi)$  are functions  of  $(\rho,\varphi),\ A<0$.
The ergosphere of the acoustic metric  is $\frac{A^2(\rho,\varphi)+B^2(\rho,\varphi))}{\rho^2}=1$  and we assume 
that it is a smooth Jordan curve.    It was proven  in [5]  that there are always black holes  for the  acoustic matrics.
In [6] we consider 
a particular  case   when  the normals  to the ergosphere are not characteristic  at any point
of the ergosphere.
In this case  there is a smooth  black hole inside the ergosphere.
When the ergosphere has characteristic points the black hole, in general,  may have corners points.  We shall study  
two typical examples.  In this section  we consider the case of finite number of characteristic points
on the ergosphere  where  the black hole is  tangent to the ergosphere,   and in the next section we shall consider  an example 
  of acoustic black hole having a corner.
  
  Let the velocity field be of the form 
  \begin{equation}													\label{eq:4.1}
  v=\frac{A}{\rho}\hat x +\frac{B(\varphi)}{\rho}\hat\theta,
  \end{equation}
where   $A<0$  is  a constant  and  $B(\varphi)$  has  finite  number of zeros
\begin{equation}													\label{eq:4.2}
B(\theta_k)=0,\ \ \ 0\leq \theta_1<\theta_2<...<\theta_p<2\pi.
\end{equation}
Here   $\rho=|A|$  is the boundary  of the black hole    and  it  touches  the ergosphere   at points  $\theta_k,1\leq k\leq p$.

Consider  the eikonal  $S(\rho,\varphi)$:
\begin{equation}                      											\label{eq:4.3}
-\eta_0+\frac{A}{\rho}S_\rho+\frac{B(\varphi)}{\rho^2}S_\varphi +\sqrt{S_\rho^2+\frac{1}{\rho^2}S_\varphi^2}=0
\end{equation}
such that $S_\rho\rightarrow +\infty$  when  $\rho\rightarrow |A|$.  As  in [6]  (cf.  also \S 3), 
for small $\rho-|A|$  it  is  convenient  to  use  
an approximation $S_1(\rho,\varphi)$ of eikonal  $S(\rho,\varphi)$  that  satisfies  the equation
\begin{equation}													\label{eq:4.4}
-\eta_0+\frac{\rho-|A|}{|A|}S_{1\rho}+\frac{B(\varphi)}{|A|^2}S_{1\varphi}=0.
\end{equation}
  We  look  for  a solution  of (\ref{eq:4.4})  in a form
  \begin{equation}													\label{eq:4.5}
  S_1(\rho,\varphi)=\eta_0|A|\ln(\rho-|A|)+S_2(\rho,\varphi),
  \end{equation}
  where $S_2(\rho,\varphi)$  satisfies
  $$
  (\rho-|A|)S_{2\rho}+\frac{B(\varphi)}{|A|}S_{2\varphi}=0,\ \ \rho>|A|.
  $$
  Let  $[\varphi_r,\varphi_{r1}]$  be  a closed  interval where  $B(\varphi)\neq 0$  and  $\varphi_{r0}\in (\varphi_r,\varphi_{r1})$.
  For  $\varphi\in [\varphi_r,\varphi_{r1}]$  we have
  \begin{equation}													\label{eq:4.6}
  S_2(\rho,\varphi)=\alpha_r\ln |\rho-A|-\alpha_r\int\limits_{\varphi_{r0}}^\varphi |A|B^{-1}(\varphi')d\varphi' + d_r,
  \end{equation}
  where  $\alpha_k$  and  $d_k$  are arbitrary.  Therefore
  \begin{equation}														\label{eq:4.7}
  S_1(\rho,\varphi)=(\eta_0|A|+\alpha_r)\ln |\rho-|A|
  +S_{3r},
  \end{equation}
  where
  \begin{equation}														\label{eq:4.8}
  S_{3r}(\varphi)=-\alpha_r\int\limits_{\varphi_{r0}}^\varphi |A| B^{-1}(\varphi')d\varphi + d_r,
  \end{equation}
  for $\varphi\in  [\varphi_r,\varphi_{r1}]$.  
  
  Let  
  $\hat C_r(x_0,\rho,\varphi)$  be a solution  of (\ref{eq:2.2})  (wave  packet)    having the following  initial conditions
  \begin{equation}														\label{eq:4.9}
 \hat C_r\Big|_{x_0=0}=\theta(\rho-|A|)c_r(\varphi)f(\rho)
  e^{i(\eta_0|A|+\alpha_r)  |\ln|\rho-|A|\ |+ S_{3r}
  }
  \end{equation}
  where
  \begin{equation}														\label{eq:4.10}
  f(\rho)=\frac{(\rho-|A|)^\e  e^{-a(\rho-|A|)}}{\sqrt \rho},
  \end{equation}
  $a>0,\e>0$  (cf.  \S 3),  $\mbox{supp\,} c_r(\varphi)\subset (\varphi_r,\varphi_{r_1})$,
   \begin{equation}														\label{eq:4.11}
  \frac{\partial \hat C_r}{\partial x_0}\Big|_{x_0=0}=
  i\beta_r
  \theta(\rho-|A|)c_r(\varphi)f(\rho)
  e^{i(\eta_0|A|+\alpha_r)|\ln|\rho-|A|\, |
  +S_{3 r}(\varphi)
  },
  \end{equation}
  \begin{multline}														\label{eq:4.12}
  \beta_r=-\frac{A}{\rho}S_{1\rho}
  -\frac{B(\varphi)}{\rho^2}S_{1\varphi}-\frac{\eta_0|A|+\alpha_r}{\rho-|A|}
  \\
  =
  -\frac{A}{\rho}\frac{(\eta_0|A|+\alpha_r)}{\rho-|A|}-\frac{(\eta_0 |A|+\alpha_r)}{\rho-|A|}
  +\frac{\alpha_k|A|}{\rho^2}
  \\
  =-\frac{\eta_0|A|+\alpha_r}{\rho} +\frac{\alpha_r|A|}{\rho^2}=-\eta_0+O(\rho-|A|).
  \end{multline}
  Note that although   (\ref{eq:4.7})  is an approximation  of the eikonal, the  solution  $C_r(x_0,\rho,\varphi)$  is  an exact  solution 
  of the wave  equation  (\ref{eq:2.2}).
  
  Finally we  define  wave packet  $\hat C(x_0,\rho,\varphi)$  as the sum
  \begin{equation}														\label{eq:4.13}
  \hat C(x_0,\rho,\varphi)=\sum_{r=1}^p \hat C_r(x_0,\rho,\varphi).
  \end{equation}
  Similarly to (\ref{eq:3.3})  the KG norm  of  $\hat C$  is 
  \begin{equation}														\label{eq:4.14}
  <\hat C,\hat C>=\sum_{r=1}^p\int\limits_0^{2\pi}|c_r(\varphi)|^2\frac{\Gamma(2\e)(\eta_0|A|+\alpha_r)}{(2a)^\e}.
  \end{equation}

  To compute the average  number of particles  created by $\hat C(x_0,\rho,\varphi)$  we use    Theorem  \ref{theo:2.1}.
  Note that 
  \begin{equation}														
  \nonumber
  \hat C^{-}(k)=-<\bar f_k^+,\hat C>=
  \sum_{r=1}^p-<\bar f_k^+, \hat C_r>=\sum_{r=1}^p  \hat C_r^{-}(k).
  \end{equation}
We have
\begin{multline}														\label{eq:4.15}
<f_k^+,\hat C_r >=i\int\limits_0^\infty\int\limits_0^{2\pi}\overline{f_k^+}
\Big(\frac{\partial \hat C_r}{\partial x_0}+\frac{A}{\rho}\frac{\partial \hat C_r}{\partial\rho}+\frac{B(\varphi)}{\rho^2}\frac{\partial\hat C_r}{\partial\varphi}\Big)
\\
-\hat C_r\Big(\frac{\partial \overline{f_k^+}}{\partial x_0}
+\frac{A}{\rho}\,\frac{\partial\overline{f_k^+}}{\partial \rho} 
 +
\frac{B(\varphi)}{\rho^2}\frac {\partial \overline{f_k^+}}{\partial \varphi}\Big)\rho d\rho d\varphi.
\end{multline}
Note that 
\begin{equation}													\label{eq:4.16}
\frac{\partial \hat C_r}{\partial\rho}=
\Big(\frac{i(\eta_0|A|+\alpha_r)}{\rho-|A|}+\frac{\e}{\rho-|A|}-a\Big)\hat C_r,
\end{equation}
\begin{equation} 													\label{eq:4.17}
\frac{\partial \hat C_r}{\partial\varphi}=i\frac{\partial S_{3r}}{\partial \varphi} \hat C_r
+\frac{\partial c_r}{\partial \varphi}\, e^{i(\eta_0|A|+\alpha_r)\ln(\rho-|A|)+iS_{3r}(\varphi)}f(\rho).
\end{equation}
Therefore we need  to take care  of the  extra turns  when  we take derivatives in $\rho$  and  in $\varphi$.
These extra terms will dissapear  when  we will take the limit when  the parameter  $a\rw\infty$.
  
  Computing  $\hat  C_r^{-}(k)$  as  in  (3.4)-(3.9)   in \S 3 we obtain
  \begin{equation}														\label{eq:4.18}
  \hat C_r^{-}(k)=\hat C_{r1}^{-}+\hat C_{r2}^{-},
  \end{equation}
  where 
  \begin{equation}														\label{eq:4.19}
\hat C_{r1}^{-}=\frac{\gamma_{m'}^{(r)}}{\sqrt 2}
\frac{\big(\eta^2_\rho+a^2\big)^{\frac{1}{4}}e^{i|A|\eta_\rho} \Gamma(i(\eta_0|A|+\alpha_r)+\e+1)
e^{i(i(\eta_0|A|+a_r)+\e+1)\frac{\pi}{2}}}
{(\eta_\rho+ia)^{i(\eta_r|A|+\alpha_r)+\e+1}},
\end{equation}
\begin{multline}														\label{eq:4.20}
\hat C_{r2}^{+-}=\frac{-\gamma_{m'}^{(r)}(\eta_0|A|+\alpha_r)}{\sqrt 2}
\\
\cdot
\ \frac{\big(\eta^2_\rho+a^2\big)^{-\frac{1}{4}}e^{i|A|\eta_\rho} \Gamma(i(\eta_0|A|+\alpha_r)+\e)
e^{i(i(\eta_0|A|+\alpha_r)+\e)\frac{\pi}{2}}}
{(\eta_\rho+ia)^{i(\eta_0|A|+\alpha_r)+\e}}
\\
+\gamma_{m'}^{(r)}O(|\eta_\rho+ia|^{-\e-1})+\hat\gamma_{m'}^{(r)}O(|\eta_\rho+ia|^{-\e-1}),
\end{multline}
where 
\begin{equation}														\label{eq:4.21}
\gamma_{m'}^{(r)}=\frac{1}{2\pi}
\int\limits_0^{2\pi}c_r(\varphi)e^{im'\varphi+
iS_{3r}(\varphi)}
 \ d\varphi,
\end{equation}
 \begin{equation}														\label{eq:4.22}
\hat\gamma_{m'}^{(r)}=\frac{1}{2\pi}
\int\limits_0^{2\pi}  iB(\varphi)\frac{\partial c_r}{\partial\varphi}
e^{im'\varphi+
iS_{3r}(\varphi)}
 \ d\varphi,
\end{equation} 
  Note  that
  \begin{multline}														\label{eq:4.23}
  \sum_{m'=-\infty}^\infty \gamma_{m'}^{(r_1)}\overline\gamma_{m'}^{(r_2)}
  =
  \frac{1}{2\pi}
  \int\limits_0^{2\pi}c_{r_1}(\varphi)
  e^{iS_{3r_1}(\varphi)}
  \overline{c_r(\varphi)}
  e^{-iS_{3r_2}(\varphi)}
  d\varphi 
   =0,
  \end{multline}
  since $c_{r_1}(\varphi)c_{r_2}(\varphi)\equiv 0.$
  Therefore  taking into account  (\ref{eq:4.23})  we  get,  as in  \S  3:
  \begin{multline}														\label{eq:4.24}
\langle 0| N(\hat C) |0\rangle=\sum_{m=-\infty}^\infty\int\limits_{-\infty}^\infty |\hat C^-(k)|^2 d\eta_\rho
=\sum_{r=1}^p\sum_{m'=-\infty}^\infty\int\limits_{-\infty}^\infty   |\hat C_r^-|^2 d\eta_\rho
\\
=\sum_{r=1}^p\frac{1}{2\pi}\int\limits_0^{2\pi}|c_r(\varphi)|^2d\varphi\ \frac{1}{2}\int\limits_{-\infty}^\infty
\big|\eta_0|A|+\alpha_r-i\e\big|^2
\Big|\frac{\eta_\rho}{(\eta_\rho^2+a^2)^{\frac{1}{4}}}
+(\eta_\rho^2+a^2)^{\frac{1}{4}}\Big|^2
\\
\cdot e^{-2\pi(\eta_0|A|+\alpha_r-i\e)}
|\Gamma_1(i(\eta_0|A|+\alpha_r)+\e)|^2e^{2(\eta_0|A|+\alpha_r)\arg(\eta_\rho+ia)}
(\eta_\rho^2+a^2)^{-\e} d\eta_\rho
\\
+\int\limits_{-\infty}^\infty O(|\eta_\rho+ia|^{-2\e-2})d\eta_\rho,
\end{multline}
where
\begin{equation}													 	\label{eq:4.25}
  \Gamma_1((\eta_0|A|+\alpha_r)+\e)=i\int\limits_0^\infty e^{i(i(\eta_0|A|+\alpha_r)+\e-1)\ln y+i(\e-1)\frac{\pi}{2}-iy}dy,
  \end{equation}
 and  we  used  that
 \begin{equation}															\label{eq:4.26}
\sum_{m'=-\infty}^\infty|\gamma_{m'}^{(r)}|^2=\frac{1}{2\pi}\int_0^{2\pi}|c_r(\varphi)|^2d\varphi,
\end{equation}
and that  (cf.  (\ref{eq:4.22}))  $|\hat\gamma_{m'}^{(r)}|\leq \frac{C}{|m'|^2}$.

Therefore we proved the following theorem.
\begin{theorem}															\label{theo:4.1}  
 The average number  of particles  created  by the  wave packet  
 (\ref{eq:4.9}),  (\ref{eq:4.11})
   is given  by the formula  (\ref{eq:4.24}).
 \end{theorem} 
  
  Finally,  making  the  change  of variables   $\eta_\rho\rw a\eta_\rho$,   replacing  $\hat C$  by the  normalized  wave  packet 
 $\hat C_n=\frac{\hat C}{<\hat C,\hat C>}$  and  taking  the limit  where  $a\rw \infty$  we get  
  \begin{multline}															\label{eq:4.27}
  \lim_{a\rw\infty}\langle 0|N(\hat C_n)|0\rangle=\Big(\sum_{r=1}^p\int\limits_0^{2\pi}|c_r(\varphi)|^2d\varphi\frac{\Gamma(2\e)}{2^\e}
  (\eta_0|A|+\alpha_r)\Big)^{-1}
  \\
  \cdot
  \sum_{r=1}^p\frac{1}{2\pi}\int\limits_0^{2\pi}|c_r(\varphi)|^2d\varphi\int\limits_{-\infty}^\infty C_{3r}(\eta_\rho)d\eta_\rho,
  \end{multline} 
   where $\hat C_{3r}(\eta_\rho)$  is the same as   $C_3$  in  (\ref{eq:3.12})   with  $|\xi_0|A|$  replaced  by  $\eta_0|A|+\alpha_r$.
   
   Note  that  when  $\alpha_r$  is  the same  for all $1\leq r\leq p$  then   the sum
   $\sum_{r=1}^p\int_0^{2\pi}|c_k(\varphi)|^2d\varphi$  in (\ref{eq:4.27})   cancels.
  
  {\bf Remark 4.1}  (cf.  Remark 3.1).  As  in the end  of  \S  3  we  have that $\int_0^\infty\hat C_{3r}(\eta_\rho)d\eta_\rho$  is exponentially
  decaying  when  $(\eta_0|A|+\alpha_r)\rw\infty$  and
  $$
  \int\limits_{-\infty}^0\hat C_{3r}(\eta_\rho)d\eta_\rho=O((\eta_0|A|+\alpha_r)^{2-\delta(1+2\e)}
  $$  
  (cf.  (\ref{eq:3.24})). 
  
\section{The Hawking radiation  in the case of black  hole with corners}
\init

In this section   
  we study the Hawking radiation from  rotating acoustic black holes  having corners.  It was proven in 
  [4],  [5]  and  [7]   that the zero energy null geodesics form  two family of smooth curves inside the ergosphere  
  and the boundary of the black hole  (the event horizon)  consists of segments  of zero energy  null geodesics 
  belonging  to one or another  family.   In the case  when  the normal  to the ergosphere  is not characteristic at any 
  point,   the black hole is formed  by one    family of null geodesics and it is smooth closed curve.  
  When  the ergosphere  contains  the characteristic  points then  the black hole (or black holes)  consists  of
  segments  belonging to the different families
  and therefore when  adjacent segments  belong  to different  families  they intersect and form  a corner.

Let,  as  in [5],  Example 4.2,  
\begin{equation}												\label{eq:5.1}
A=A_0+\e  r\sin \varphi,\ \ B=\e  r\cos\varphi,  \ \ 0< \e < 1,\ A_0< -1 ,
\end{equation}
$A$  and  $B$   define
 the velocity field  by formula   (\ref{eq:2.1}).

The  equation  of the ergosphere is
\begin{equation}												\label{eq:5.2}
r=r_0(\varphi)=-\frac{A_0}{1-\e^2}\Big(-\e\sin\varphi +\sqrt{1-\e^2\cos^2\varphi}\ \Big).
\end{equation}
Points  $\alpha_1=(r_0(\frac{\pi}{2}),\frac{\pi}{2})$  and $\alpha_2=(r_0(-\frac{\pi}{2}),-\frac{\pi}{2})$
on the ergosphere  are characteristic points.  Note that both families   of  zero-energy  null-geodesics  are  tangent  to the ergosphere
at the characteristic point.  
For the ergosphere   (\ref{eq:5.2})  we have zero energy  null-geodesics     $\gamma_1$   and  $\gamma_2$  that starts    at  
$\alpha_1=(r_0(\frac{\pi}{2}),\frac{\pi}{2})$  and  intersected  at 
some point   $\alpha_3$  forming an angle   (see Fig. 1)

\begin{tikzpicture}
[scale=0.8]

\draw[ultra thick](-3.2,0).. controls (-3,2)  and (-2,3) .. (0,3.24);
\draw[ultra thick](-3.2,0).. controls (-3,-2)  and (-2,-3) .. (0,-3.2);

\draw[ultra thick](3.2,0).. controls (3,2)  and (2,3) .. (0,3.24);
\draw[ultra thick](3.2,0).. controls (3,-2)  and (2,-3) .. (0,-3.2);

\draw[ultra thick](-1,-3).. controls (1,-2)  and (2,2) .. (0,3.24);

\draw[ultra thick](1,-3).. controls (-1,-2)  and (-2,2) .. (0,3.24);

\draw (1.4,1.) node {$\gamma_2$};
\draw (-1.4,1.) node {$\gamma_1$};

\draw (0.3,3.5) node {$\alpha_1$};
\draw (0.3,-3.5) node {$\alpha_2$};
\draw (0.45,-2.2) node {$\alpha_3$};

\draw (0,3.22) circle  (2.7pt);
\draw (0,-3.2) circle  (2.5 pt);
\draw (0,-2.18) circle  (2.7 pt);
\end{tikzpicture}
\\
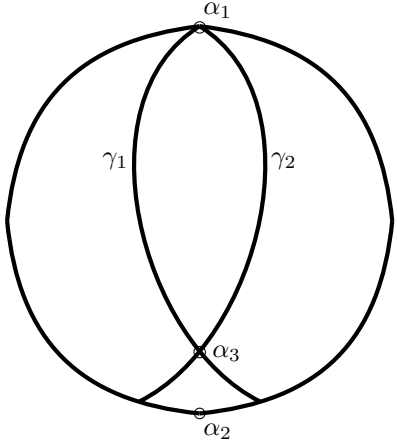
{\bf Fig. 1.} Null-geodesics  $\gamma_1$  and  $\gamma_2$  intersect  at 
point $\alpha_3$  
and   bound a black hole

\ \ \ \ \ \ \ \ having corner  point $\alpha_3$.

Denote   by  $\gamma_{10}$  the  part  of  $\gamma_1$  where  $   \varphi_1<\varphi<\varphi_{11}$   on  $\gamma_1$.
Analogously let  $\gamma_{20}$  be the   part  of  $\gamma_2$  such  that  $\varphi_2<\varphi<\varphi_{20}$  on  $\gamma_2$.
Let  $\rho=\rho_i(\varphi)$  be the equation   of   $\gamma_{i0},\ \varphi_i<\varphi <\varphi_{i1},i=1,2$.

Connect  $\rho_1(\varphi_1)$  and  $\rho_2(\varphi_{21})$  and  connect $\rho_1(\varphi_{11})$  and  $\rho_2(\varphi_2)$ by smooth
curves  to  get  a smooth periodic  curve $\rho=\rho_0(\varphi),\ 0<\varphi < 2\pi$    
such  that   $\rho_0(\varphi)=\rho_i(\varphi)$  for   $\varphi_1\leq  \varphi\leq  \varphi_{11}$  and  
$\rho_0(\varphi)=\rho_2(\varphi)$  for   $\varphi$  between  $\varphi_{21}$  and  $\varphi_2$.

The following arguments are not restricted    to  Example   4.2  and apply  to any  situation 
when we have two smooth segments  $\gamma_{10}$  and  $\gamma_{20}$    of the boundary of the black hole.

As in  [6], \S 4.2,
make  change of variable
\begin{align}																			\label{eq:5.3}
&\tilde \rho=\rho-\rho_0(\varphi),
\\
\nonumber
&\varphi=\varphi.
\end{align}
Denote by    $\tilde f_k^+,  k=(\eta_\rho,m)$,    the solutions  of  the wave  equation  in  $(x_0,\tilde\rho,\varphi)$  coordinates such that
\begin{equation}																	\label{eq:5.4}
\tilde f_k^+(x_0,x)\Big |_{x_0=0}=
\tilde\gamma_k e^{i\eta_{\tilde \rho}\tilde \rho +im'\varphi},
\end{equation}
\begin{equation}																	\label{eq:5.5}
\frac{\partial f_k^+}{\partial x_0}\Big|_{x_0=0}=i\tilde \lambda _0^- (k)\tilde\gamma_k e^{i\eta_{\tilde \rho}\tilde\rho +im'\varphi},
\end{equation}
where
\begin{equation}																		\label{eq:5.6}
\tilde\gamma_k=\frac{1}{\sqrt{\rho_0(\varphi)+\tilde\rho}\,\,\big(\eta_{\tilde\rho}^2+a^2\big)^{\frac{1}{4}}\sqrt{2(2\pi)^2}},\ \ 
\rho=\rho_0(\varphi)+\tilde\rho.
\end{equation}
\begin{equation}																		\label{eq:5.7}
\tilde\lambda_0^-(k)=-\Big(\frac{A}{\rho}-\frac{B}{\rho^2}\rho_0'(\varphi)\Big)\eta_{\tilde \rho}
 -\frac{B}{\rho^2}m'  -\sqrt{\eta_{\tilde \rho}^2+a^2},
\end{equation}
The eikonal  equation  in $(\tilde\rho,\varphi)$  coordinates  takes the  form (cf. (4.19)   in [6])
\begin{equation}
\nonumber																		
\tilde A(\tilde\rho,\varphi)\tilde S_{\tilde\rho}^2 +2\tilde B(\eta_0,\tilde\rho,\varphi,\tilde S_\varphi)\tilde S_{\tilde\rho}
+\tilde C(\eta_0,\tilde\rho,\varphi,\tilde S_\varphi)=0,
\end{equation}
where 
$\tilde A,\tilde B,\tilde C$  are the same  as  in  (4.20)  in [6].  Since  
$\tilde \rho=0$  is characteristic  when  $\varphi_1<\varphi <\varphi_{11}$  and when  $\varphi_2<\varphi<\varphi_{21}$  we  have that
$\tilde A(\tilde \rho,\varphi)=A_{01}(\tilde\rho,\varphi)\tilde\rho$  when $\varphi_1<\varphi <\varphi_{11}$  and when  $\varphi_2<\varphi<\varphi_{21}$.

As in [6]  we have that  
\begin{equation}																		\label{eq:5.8}
\tilde\rho S_{\tilde\rho}^{(j)}(\tilde\rho,\varphi)+B_1^{(j)}(\varphi)\eta_0+B_2^{(j)}S_\varphi^{(j)}=0,
\end{equation}
where
equation  (\ref{eq:5.8})    for  $j=1$  holds   on  $(\varphi_1,\varphi_{11})$  and  (\ref{eq:5.8})  for $j=2$  holds
on  $(\varphi_2,\varphi_{21})$.

Similarly  to  \S 4   the solution  of  (\ref{eq:5.8})  has the form
\begin{equation}                        															\label{eq:5.9}
S^{(j)}(\tilde\rho,\varphi)=\tilde\alpha_j\ln \tilde\rho 
-\int\limits_{\varphi_{j0}}^\varphi\frac{\tilde \alpha_j+ B_1^{(j)}(\varphi')\eta_0}{B_2^{(j)}}d\varphi'+\tilde d_j,
\end{equation}
where  $\tilde\alpha_j>0$   and  $d_j$  are arbitrary,   $\varphi_{j0}\in (\varphi_j,\varphi_{j1})$.

Note that  (\ref{eq:5.9})  for $j=1,2$  holds  when  $\varphi_j<\varphi<\varphi_{j1}$.

Now we shall  construct  the wave  packet  $\tilde C(x_0,\tilde \rho,\varphi)$   in the form
\begin{equation}																\label{eq:5.10}
\tilde C=\tilde  C_1+\tilde C_2,
\end{equation}
where 
\begin{equation}		 														\label{eq:5.11}
\tilde C_j\Big|_{x_0=0}=s_j(\varphi)\theta(\tilde \rho) e^{-i( \tilde\alpha_j\ln \tilde\rho 
-\int\limits_{\varphi_{j0}}^\varphi\frac{\tilde \alpha_j+\tilde B_1^{(j)}(\varphi')\eta_0}{B_2^{(j)}(\varphi')}d\varphi'+  \tilde d_j)}\ 
\frac{e^{-a\tilde\rho}\tilde \rho^\e}{\sqrt{\rho_i(\varphi)+\tilde\rho}},
\end{equation}
\begin{equation}																\label{eq:5.12}
\frac{\partial\tilde C_j}{\partial  x_0}\Big|_{x_0=0}=i\tilde\beta_j\tilde C_j\Big|_{x_0=0},
\end{equation}
where
\begin{equation}																\label{eq:5.13}
\tilde\beta_j^{i}=\Big(\frac{A}{\rho}-\frac{B}{\rho^2}\rho_0'(\varphi)\Big)\frac{\tilde\alpha_j}{\tilde  \rho}-
\frac{B}{\rho^2}\frac{\partial S_j}{\partial\varphi}-\frac{\tilde\alpha_j}{\rho},\ \varphi\in (\varphi_j,\varphi_{j1}),
\end{equation},
\begin{equation}   																\label{eq:5.14}
\mbox{supp\ } s_j(\varphi) \subset (\varphi_j,\varphi_{j1}),\ j=1,2.
\end{equation}
Denote 
$$
\tilde \gamma_m^{(j)}=\frac{1}{2\pi}\int\limits_0^{2\pi}s_j(\varphi)
 e^{-i\big(  -\int\limits_{\varphi_{j0}}^\varphi\frac{\tilde \alpha_j+ B_1^{(j)}(\varphi')\eta_0}{B_2^{(j)}}
 d\varphi'+\tilde d_j+m\varphi\big)} d\varphi.
$$
By the Parseval's  equality
\begin{equation}																\label{eq:5.15}
\sum_{m=-\infty}^\infty|\tilde\gamma_m^{(j)}|^2=\frac{1}{2\pi}\int\limits_0^{2\pi}|s_j(\varphi)|^2d\varphi.
\end{equation}
Therefore as  in  \S 3 and \S 4 computing  the average  number  of created particles 
we get
\begin{multline}																			\label{eq:5.16}
\langle 0 | N(\tilde C) | 0\rangle
 =
\sum_{j=1}^{2}
\frac{1}{2\pi}\int_0^{2\pi}|s_j(\varphi)|^2d\varphi
\\
\cdot \frac{1}{2}\int\limits_{-\infty}^\infty |\tilde\alpha_j-i\e|^2\Big|\frac{\eta_\rho}{(\eta_\rho^2+a^2)^{\frac{1}{4}}}
+(\eta_\rho^2+a^2)^{\frac{1}{4}}\Big|^2 
\\
\cdot  e^{-2\pi\tilde\alpha_j}
|\Gamma_1(i\tilde\alpha_j+\e)|^2e^{2 \tilde\alpha_j \arg(\eta_\rho+ia)}(\eta_\rho^2+a^2)^{-\e-1}d\eta_\rho
\\
+\int\limits_{-\infty}^\infty O(|\eta_\rho+ia|^{-2\e-2})d\eta_\rho.
\end{multline}
Thus we proved  the following theorem:
\begin{theorem}																		\label{theo:5.1}
The average  number  of particles  created  by the wave  packet  (\ref{eq:5.11}),   (\ref{eq:5.12})  is given  by (\ref{eq:5.16}).
\end{theorem}
As   in \S 4,  replacing  $\tilde C$  by   the normalized  wave  packet $\tilde C_n$  and  taking  the  limit  as  $a\rw\infty$   we get  that
$\lim_{a\rw\infty}\langle 0|N(\tilde C_n|0\rangle$  has an expression  similar  to  
(\ref{eq:4.27})  with   $\eta_0|A|+\alpha_r$  replaced  by  $\tilde \alpha_r$.

\end{document}